\documentstyle[preprint,aps,psfig]{revtex}
\tighten
\begin{document}

\title{Evidence for a missing nucleon resonance in kaon photoproduction}
\author{T. Mart}
\address{Jurusan Fisika, FMIPA, Universitas Indonesia, Depok 16424, Indonesia}
\author{C. Bennhold}
\address{Center for Nuclear Studies, Department of Physics, The George 
         Washington University, Washington, D.C. 20052, USA}

\date{\today}
\maketitle

\begin{abstract}
New {\small SAPHIR} $p(\gamma,K^+)\Lambda$ total cross section data show a
resonance structure at a total c.m. energy around 1900 MeV. We investigate 
this feature with an isobar model and find that the structure
can be well explained by including a new $D_{13}$ resonance at 1895 MeV.
Such a state has been predicted by a relativistic quark model 
at 1960 MeV with significant $\gamma N$ and $K \Lambda$ branching ratios.
We demonstrate how the measurement of the photon asymmetry
can be used to further study this resonance. In addition, 
verification of the predicted large decay
widths into the $\eta N$ and $\eta' N$ channels would allow distinguishing 
between other nearby $D_{13}$ states.\\ \\
PACS number(s): 14.20.Gk, 25.20.Lj, 13.60.Le, 13.30.Eg 
\end{abstract}

\newpage
\vspace{1cm}

The physics of nucleon resonance excitation continues to provide 
a major challenge to hadronic physics \cite{nstar} 
due to the nonperturbative nature of QCD
at these energies.  While methods like Chiral Perturbation Theory are not
amenable to $N^*$ physics, lattice QCD has only recently begun to contribute
to this field.  In a recent study \cite{lee98} the excitation energies of 
$1/2^-$ and $3/2^-$ baryon resonances are calculated for the first time 
on the lattice with improved actions. The results show a clear splitting 
of these states from the ground state nucleon, demonstrating
the potential and the promise of extracting
$N^*$ structure from lattice QCD.  However, most of the theoretical work
on the nucleon excitation spectrum has been performed in the realm of
quark models. Models that contain three constituent valence quarks
predict a much richer resonance spectrum \cite{NRQM,capstick94}
than has been observed in $\pi N\to \pi N$ scattering experiments. 
Quark model studies have suggested that those "missing" resonances 
may couple strongly to other channels, 
such as the $K \Lambda$ and $K \Sigma$ channels \cite{capstick98}
or final states involving vector mesons.

\begin{figure}[htbp]
  \begin{center}
    \leavevmode
    \psfig{figure=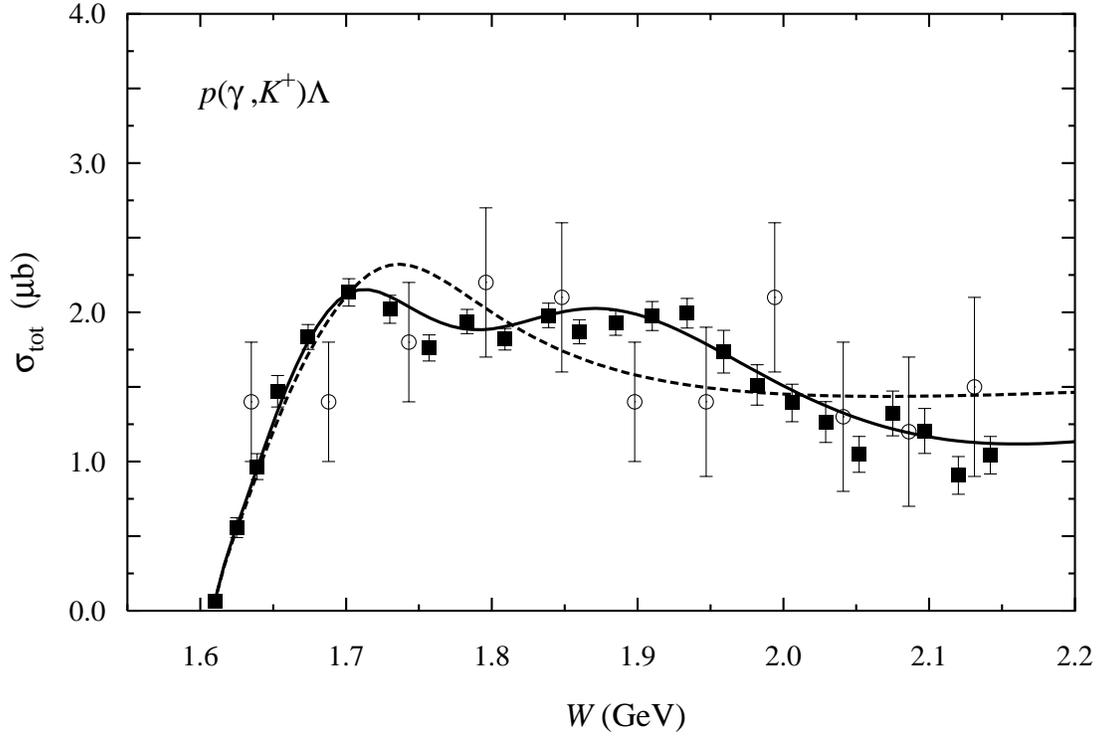,width=140mm}
    \caption{Total cross section for $K^+\Lambda$ photoproduction
        on the proton. The dashed line shows the model without the 
      	$D_{13}(1960)$ resonance, while the solid line is obtained 
	by including the $D_{13}(1960)$ state. The new {\small SAPHIR} 
	data \protect\cite{saphir98} are denoted by the solid 
      	squares, old data \protect\cite{old_data} are shown by the 
	open circles.}
   \label{fig:totalcs}
  \end{center}
\end{figure}

The newly established electron and photon facilities have made 
it possible to investigate the mechanism of nucleon resonance excitation
with photons with much improved experimental accuracy. 
Experiments with kaon-hyperon final states have been performed at 
{\small ELSA} \cite{saphir98} and are being analyzed at JLab. 
Much improved data are becoming available in the $p(\gamma, K^+)\Lambda$,
$p(\gamma, K^+)\Sigma^0$ and $p(\gamma, K^0)\Sigma^+$ channels,
from total cross section to polarization observables. 
The new {\small SAPHIR} total cross section data \cite{saphir98} for 
the $p(\gamma, K^+)\Lambda$ channel, shown in Fig.~\ref{fig:totalcs},
indicate for the first time a 
structure around $W = 1900$ MeV. This structure could not be resolved before, 
due to the low quality of the old data. It is the purpose of this 
paper to investigate this structure in the framework of an isobar model.

Pioneered by Thom \cite{thom66}, most studies over the last 30 years 
analyzed the $N(\gamma, K)\Lambda (\Sigma) $ in a tree-level isobar 
framework \cite{adelseck85,williams92,mart95,saghai96}
that included a number of resonances whose couplings were adjusted to
reproduce the experimental data.  Due to the poor data quality it was 
not possible to decide which resonances contributed, even the magnitude
of the background terms was uncertain. Recently, two new developments 
have provided significant progress in this field. First, a coupled-channels
calculation that included final-state interactions \cite{feuster98} linked
the photoproduction process $p(\gamma, K^+)\Lambda$ to the hadronic process
$p(\pi^-, K^0)\Lambda$. Secondly, the recent work on including hadronic form 
factors in photoproduction reactions \cite{haberz,hbmf98} while maintaining 
gauge invariance has resulted in the proper description
of the background terms, allowing the use of approximate SU(3) symmetry to 
fix the Born coupling constants $g_{K \Lambda N}$ and $g_{K \Sigma N}$.

Due to their isospin structure the $K \Sigma$ photoproduction 
channels can involve the excitation of $N^*$ as well as $\Delta$ states. 
On the other hand, $K \Lambda$ photoproduction only involves intermediate 
isospin 1/2 resonances and is therefore easier to describe.  Here, we use 
the tree-level isobar model described in Ref.\,\cite{elba98} to analyze the
$p(\gamma, K^+)\Lambda$ process in more detail.  
Guided by a recent coupled-channels analysis 
\cite{feuster98}, the low-energy resonance part of
this model includes three states that have been found to have 
significant decay widths into the $K^+\Lambda$ channel, 
the $S_{11}$(1650), $P_{11}$(1710), and $P_{13}(1720)$ resonances. 
In order to approximately account for unitarity corrections at tree-level 
we include energy-dependent widths along with partial branching fractions 
in the resonance propagators \cite{elba98}. The background part includes 
the standard Born terms along with the $K^*$(892) and $K_1$(1270) 
vector meson poles in the $t$-channel. As in Ref.\,\cite{elba98}, we 
employ the gauge method of Haberzettl \cite{haberz,hbmf98}
to include hadronic form factors.
The fit to the data was significantly improved by
allowing for separate cut-offs for the background and resonant sector.
For the former, the fits produce a soft value around 800 MeV,
leading to a strong suppression of the background terms while 
the resonant cut-off is determined to be 1900 MeV.  

As shown in Fig.~\ref{fig:totalcs}, our model cannot reproduce
the {\small SAPHIR} total cross section data
 without inclusion of a new resonance with a mass of
around 1900 MeV. While there are no 3- or 4-star isospin 1/2 
resonances around 1900 MeV in the Particle Data Table \cite{pdg98}, 
several 2-star states are listed. 
Of those only the $D_{13}(2080)$ state has been identified in older 
$p(\pi^-, K^0)\Lambda$ analyses \cite{saxon,bell} to have a noticeable 
branching ratio into the $K \Lambda$ channel.  On the theoretical side, 
the constituent quark model by Capstick and 
Roberts \cite{capstick94} predicts many new states around 1900 MeV; 
however, only a few of them have been calculated to have
a significant $K \Lambda$ decay width \cite{capstick98}.
These are the $[S_{11}]_3$(1945), $[P_{11}]_5$(1975), 
$[P_{13}]_4$(1950), and $[D_{13}]_3$(1960) states,
 where the subscript refers to the particular
band that the state is predicted in.  We have performed fits for each 
of these possible states, allowing the fit to determine the mass, width 
and coupling constants  of the resonance. We found that all four states 
can reproduce the structure at $W$ around 1900 MeV, reducing the $\chi^2/N$ 
from around 4.5 to around 3 in each case. Table \ref{table_cc2} compares 
our extracted resonance parameters with the quark model predictions of 
Ref.\,\cite{capstick98}.  While all four of the above resonances have 
large decay widths into the $K \Lambda$ channel, only the  $D_{13}$(1960) 
state is predicted to also have significant photocouplings. Table 
\ref{table_cc2} presents the remarkable agreement, up to the sign, between 
the quark model prediction and our extracted results for the $D_{13}$(1960).
The sign remains ambiguous, since at this stage we only extract the 
product of coupling constants.
For the other three states the partial widths extracted from our fit 
overestimate the quark model results by up to a factor of 30.  

How reliable are the quark model predictions? Clearly, one test is to confront
its predictions with the extracted couplings for the well-established
resonances in the low-energy
regime of the $p(\gamma, K^+)\Lambda$ reaction,
the $S_{11}(1650)$, $P_{11}(1710)$ and $P_{13}(1720)$ excitations.
Table \ref{table_cc1} shows that  the magnitudes of the 
extracted partial widths for the  
$S_{11}(1650)$, $P_{11}(1710)$, and $P_{13}(1720)$ are in good agreement 
with the quark model. Therefore, even though the remarkable quantitative 
agreement in the case of the 
$D_{13}$(1960) is probably fortuitous, we believe the structure in the
{\small SAPHIR} data is in all likelihood produced by this particular 
resonance. Is this state identical to the 2-star resonance $D_{13}$(2080) 
listed in the Particle Data Table? Table \ref{table_d13} displays a list of
$D_{13}$ states below 2.2 GeV predicted by 
Refs.\,\cite{capstick94,capstick98}, along with the Particle Data Table
listings. A closer examination of the literature reveals
that there is some evidence for two resonances in this wave
between 1800 and 2200 MeV \cite{cutkosky80}; one with a mass centered around 
1900 MeV and another with mass around 2080 MeV. It is the former
which has been seen prominently in two separate $p(\pi^-, K^0)\Lambda$
analyses \cite{saxon,bell}. Thus, we believe that the state
appearing in the {\small SAPHIR} data is in fact identical to the one seen
in hadronic $K \Lambda$ production and corresponds to the
$D_{13}$(1960) state predicted by the quark model.  The $D_{13}$
excitation around 2080 MeV seen in Refs.\,\cite{cutkosky80,hoehler79} may 
well correspond to the quark model state $D_{13}$(2055) in the $N=4$ band.
In order to clearly separate these nearby $D_{13}$ states, measuring other
channels will be helpful. For example, Ref.\,\cite{capstick94} predicts the
$D_{13}(1960)$ to have large decay widths into the $\eta N$ and
$\eta 'N$ channels, in contrast to the $D_{13}(2055)$ whose branching 
ratios into these channels are negligible.

\begin{figure}[htbp]
  \begin{center}
    \leavevmode
    \psfig{figure=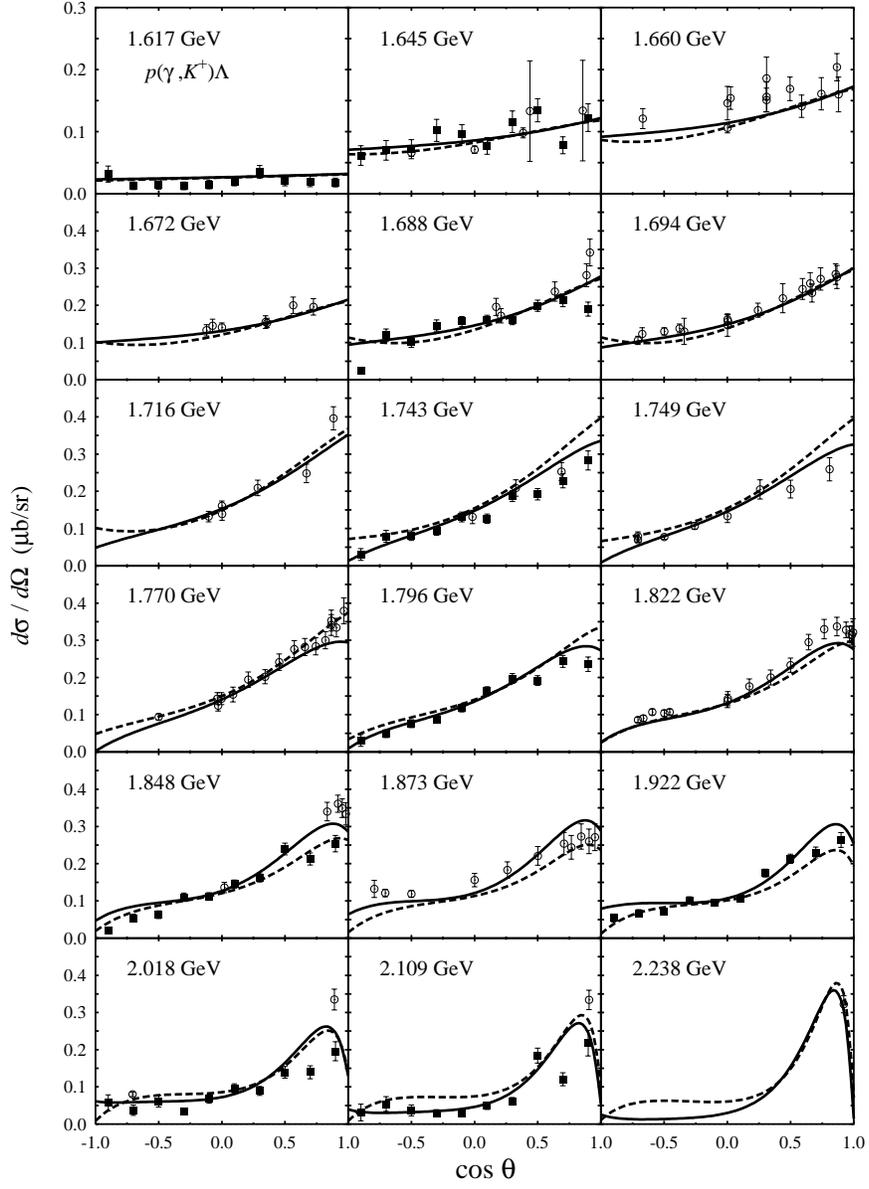,width=120mm}
    \caption{Same as in Fig. \protect\ref{fig:totalcs} for the differential 
	cross section. The total c.m. energy $W$ is 
      shown in every panel.}
    \label{fig:difcs}
  \end{center}
\end{figure}

\begin{figure}[htbp]
  \begin{center}
    \leavevmode
    \psfig{figure=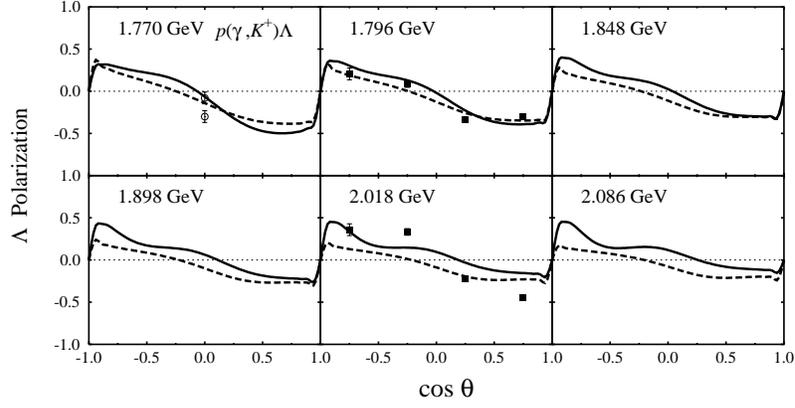,width=110mm}
    \caption{Same as in Fig. \protect\ref{fig:totalcs} for the $\Lambda$  
             recoil polarization. The total c.m. energy $W$ is 
             shown in every panel.}
    \label{fig:lambdapol}
  \end{center}
\end{figure}

Figure \ref{fig:totalcs} compares our models with and without the
$D_{13}$(1960) with the {\small SAPHIR} total cross section data.
Our result without this resonance shows only one peak
near threshold, while inclusion of the new resonance leads
to a second peak at $W$ slightly below 1900 MeV,
in accordance with the new {\small SAPHIR} data.
The difference between the two calculations is much smaller for 
the differential cross sections, as displayed in 
Fig. \ref{fig:difcs}. As expected, including the $D_{13}$(1960)
does not affect the threshold and low-energy regime
while it does improve the agreement at higher energies.
Figure \ref{fig:lambdapol} compares the recoil polarization for the two
calculations. Clearly, the differences are small for all angles,
demonstrating that the recoil polarization is not the appropriate
observable to further study this resonance.

\begin{figure}[htbp]
  \begin{center}
    \leavevmode
    \psfig{figure=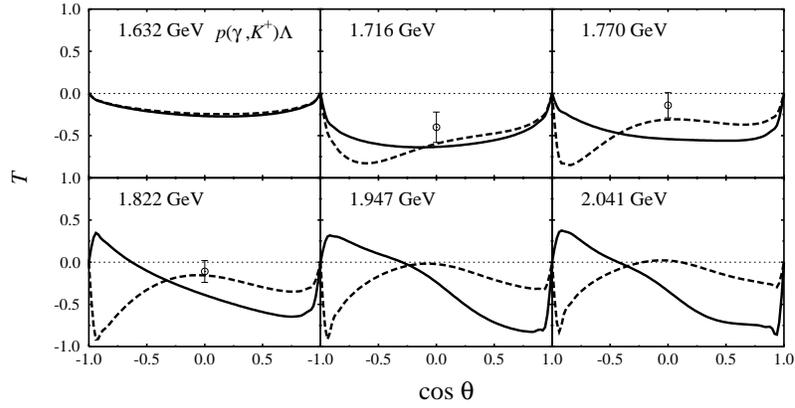,width=110mm}
    \caption{Same as in Fig. \protect\ref{fig:totalcs} for the target 
             polarization. The total c.m. energy $W$ is shown in every panel.}
    \label{fig:target}
  \end{center}
\end{figure}

\begin{figure}[htbp]
  \begin{center}
    \leavevmode
    \psfig{figure=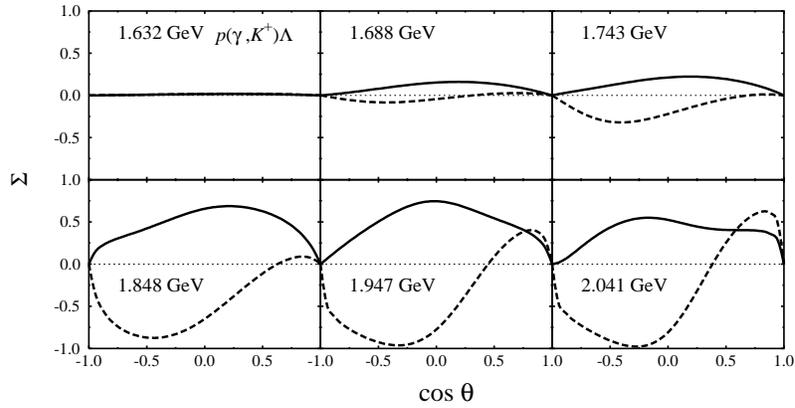,width=110mm}
    \caption{Same as in Fig. \protect\ref{fig:totalcs} for the photon 
	     asymmetry. The total c.m. energy $W$ is shown in every panel.}
    \label{fig:asymmetry}
  \end{center}
\end{figure}

The target asymmetry of $K^+\Lambda$ photoproduction is shown in 
Fig. \ref{fig:target}. Here we find larger variations between the two
calculations, especially for higher
energies. The three data points seem to favor a model without the new
$D_{13}(1960)$; however, more complete and accurate measurements 
are clearly needed over the whole angular range before any conclusion 
can be drawn. The largest effects are found in the photon asymmetry 
shown in Fig. \ref{fig:asymmetry}. For $W\geq 1800$
MeV, including the new resonance leads to a sign change in the photon 
asymmetry whose magnitude is almost one at intermediate angles.
Therefore, we would suggest that measuring this observable is 
well suited to shed more light on the contribution of this state 
in kaon photoproduction.

In conclusion, we have investigated the structure around $W= 1900$ MeV
in the new {\small SAPHIR} total cross section data in 
the framework of an isobar model.
We found that the data can be well reproduced by including a new
$D_{13}$ resonance with a mass, width and coupling parameters in good 
agreement with the values predicted by a recent quark model calculation.
To further elucidate the role and nature of this state we suggest
measurements of the polarized photon asymmetry around $W = 1900$ MeV
for the $p(\gamma, K^+)\Lambda$ reaction.  With the arrival of
new, high-precision cross section and polarization data the kaon 
photoproduction process will be able to 
unfold its full potential in the search and study of nucleon resonances.

\vspace{5mm}
 
TM thanks the member of the Center for Nuclear Studies for the 
hospitality extended to him during his stay in Washington, D.C.
This work was supported by the University Research for Graduate 
Education (URGE) grant (TM), and US DOE grant DE-FG02-95ER-40907 (CB).

\begin{table}[htbp]
\caption{Comparison between the results from our fit to the kaon 
  photoproduction data $p(\gamma, K^+)\Lambda$ (Fit) and those 
  of the quark model (QM), where the QM photocouplings were taken 
  from Ref.\,\protect\cite{capstick92}
  and the $K \Lambda$ decay widths from Ref.\,\protect\cite{capstick98}.}
\renewcommand{\arraystretch}{1.4}
\label{table_cc2}
\begin{tabular}{llrrr}
Missing Resonance & Model & $m_{N^*}$ & $\Gamma_{N^*}$ &
$\sqrt{\Gamma_{N^*N\gamma}\Gamma_{N^*K\Lambda}}/\Gamma_{N^*}$ \\
 & & (MeV) & (MeV) & ($10^{-3}$) \\ [0.5ex]
\tableline
$S_{11}(1945)$ & Fit & $1847$ & $258$ & $-10.370\pm 0.875$ \\
               & QM  & $1945$ & $595$ & $0.298\pm 0.349$ \\ [2ex]
$P_{11}(1975)$ & Fit & $1935$ & $131$ & $9.623\pm 0.789$ \\
               & QM  & $1975$ & $ 45$ & $1.960\pm 0.535$ \\ [2ex]
$D_{13}(1960)$ & Fit & ${\bf 1895}$ & ${\bf 372}$ &
        ${\bf 2.292^{+0.722}_{-0.204}}$ \\
               & QM  & ${\bf 1960}$ & ${\bf 535}$ &
        ${\bf -2.722\pm 0.729}$ \\ [2ex]
$P_{13}(1950)$ & Fit & $1853$ & $189$ & $1.097^{+0.011}_{-0.010}$ \\
               & QM  & $1950$ & $140$ & $-0.334\pm 0.070$\\ [0.5ex]
\end{tabular}
\end{table}

\begin{table}[htbp]
\caption{Comparison between the extracted fractional decay widths and
        the result from the quark model \protect\cite{capstick98,capstick92} 
	for the $S_{11}(1650)$, $P_{11}(1710)$, and $P_{13}(1720)$ 
	resonances.}
\renewcommand{\arraystretch}{1.4}
\label{table_cc1}
\begin{tabular}{lcc}
&\multicolumn{2}{c}{$\sqrt{\Gamma_{N^*N\gamma}
\Gamma_{N^*K\Lambda}}/\Gamma_{N^*}$ ($10^{-3}$)} \\
\cline{2-3} Resonance & Extracted & Quark Model \\ [0.5ex]
\tableline
$S_{11}(1650)$ & $-4.826\pm 0.051$ & $-4.264\pm 0.984$ \\
$P_{11}(1710)$ & $ ~~1.029\pm 0.172$ & $-0.535\pm 0.115$ \\
$P_{13}(1720)$ & $~ 1.165^{+0.041}_{-0.039} $ & $-1.291\pm 0.240$\\ [0.5ex]
\end{tabular}
\end{table}

\begin{table}[htbp]
\caption{Summary of listed $D_{13}$ resonances. The observed states from the 
        Particle Data Table are ordered according to 
        Refs.\,\protect\cite{capstick94,capstick98}.}
\renewcommand{\arraystretch}{1.4}
\label{table_d13}
\begin{tabular}{ccc}
Quark Model \protect\cite{capstick94,capstick98} & \multicolumn{2}{c}{Particle
Data Table \protect\cite{pdg98}}\\ \cline{2-3} Name & Name & Status \\ [0.5ex]
\tableline
$[N\frac{3}{2}^-]_1 (1495)$ & $N(1520)D_{13}$&${****}$  \\
$[N\frac{3}{2}^-]_2 (1625)$ & $N(1700)D_{13}$&${***}$ \\
$[N\frac{3}{2}^-]_3 (1960)$ & $N(2080)D_{13}$&${**}$ \\
$[N\frac{3}{2}^-]_4 (2055)$ & - &- \\
$[N\frac{3}{2}^-]_5 (2095)$ & - &- \\
$[N\frac{3}{2}^-]_6 (2165)$ & - &- \\
$[N\frac{3}{2}^-]_7 (2180)$ & - &- \\
[0.5ex]
\end{tabular}
\end{table}

\end{document}